\documentclass{agujournal2018}
\usepackage{apacite}
\usepackage{url} 

\usepackage{graphicx}
\usepackage{amsmath}
\usepackage[varg]{txfonts}
\usepackage{xspace}
\usepackage{siunitx}
\usepackage[version=3]{mhchem}
\usepackage[utf8]{inputenc}
\usepackage{hyperref}

\newcommand{\NtwoB}{\ensuremath{\mathrm{N}_2(B^3\Pi_g)}\xspace}
\newcommand{\NtwoC}{\ensuremath{\mathrm{N}_2(C^3\Pi_u)}\xspace}

\draftfalse

\journalname{Geophysical Research Letters}

\begin{document}

\title{Spontaneous emergence of space stems ahead of negative leaders in lightning and long sparks}

\authors{A. Malagón-Romero\affil{1}, A. Luque\affil{1}}

\affiliation{1}{IAA-CSIC, P.O. Box 3004, 18080, Granada, Spain}

\correspondingauthor{A. Malagón-Romero}{amaro@iaa.es}

\begin{keypoints}
\item Space stems start as regions of locally lower conductivity in the streamer channels around the tip of negative leaders. 
\item An attachment instability enhances the electric field at the space stem decreasing further the conductivity and leading to bright and locally warmer regions.
\item As the attachment instability is favoured by high electric fields inside negative streamer channels it explains why only negative leaders propagate in long steps.
\end{keypoints}

\begin{abstract}
  We investigate the emergence of space stems ahead of negative leaders.
  These are luminous spots that appear ahead of an advancing
  leader mediating the leader's stepped propagation.
  We show that space stems start as regions of locally depleted conductivity
  that form in the streamers of the corona
  around the leader.  An attachment instability enhances the electric field
  leading to strongly inhomogeneous, bright and locally warmer regions 
 ahead of the leader that explain the existing observations.  Since the attachment
  instability is only triggered by fields above 
\SI{10}{kV/cm} and internal electric fields are lower in
  positive than in negative streamers, our results explain why, although
  common in negative leaders, space stems and stepping
  are hardly observed if not absent in positive leaders.
   Further work is required to fully explain the
  streamer to leader transition, which requires an electric current persisting
  for timescales longer than the typical attachment time of electrons, around
  \SI{100}{ns}.

\end{abstract}

\section*{Plain Language Summary}
\noindent
Long electrical discharges of negative polarity, such as most cloud-to-ground lightning flashes, propagate in a stepped manner, i.e. alternating between standing and jumping suddenly. The underlying mechanism explaining this behavior is not well understood, although we know that space stems are a key element. These are bright and locally warmer segments that appear ahead of a discharge channel and apparently isolated from it. For the first time, we show how these space stems emerge spontaneously in our simulations from regions of locally lower conductivity that latter become bright and warm. Then on one hand we propose a possible origin of the space stems and, on the other hand, we shed some light on possible mechanisms that grow these stems to longer times, beyond 100 ns.

\section{Introduction}
One of the outstanding mysteries in atmospheric electricity concerns the progression of negative lightning leaders.  Being hot and ionized channels, leaders are initiated in a thundercloud and expand bipolarly, with their positive and negative extremes advancing in more or less opposite directions.  For some elusive reason, negative leaders advance in a stepped fashion, with waiting times of tens of microseconds punctuated by sudden jumps of microsecond timescale \citep{Dwyer2014/PhR}.  This behavior is observed not only in lightning leaders but also in negative laboratory discharges longer than about two meters.

Besides being a fundamental but mysterious process in electrical discharges,
leader steps are relevant because they produce the Very High Frequency (VHF) radio pulses that reveal the development of lightning flashes in Lightning Mapping Arrays \citep{Thomas2001/GeoRL}.  Leader steps are also correlated with X-ray emissions detected around a lighting discharge \citep{Dwyer2005/GeoRL/3} and therefore they are possibly linked to Terrestrial Gamma-ray Flashes (TGFs) detected by satellites orbiting hundreds of kilometers above ground \citep{Fishman1994/Sci,Smith2005/Sci,Marisaldi2010/PhRvL,Briggs2013/JGRA}.

The first observation of leader stepping can be traced back to the
pioneering work of \citet{Schonland1935/RSPSA} in the 1930s, who coined the term ``stepped leader'' 
for the intermittent advance of downward negative lightning channels 
recorded in their streak camera.  In the decades after Schonland's work,
advances in this topic arose mostly from laboratory experiments with meter-long spark discharges.  The work of, among others, \citet{Gorin1976/IEEConf} and the \citet{LesRenardiers1978/Elektra} revealed the dynamics of a negative leader step: the leader tip is preceded by a filamentary corona containing a bright nucleus termed ``space stem''.  After some microseconds the space stem evolves into a ``space leader'' that propagates in both directions and whose extremes are surrounded by 
additional coronas of both polarities.  The leader completes one step when the space leader bridges the gap to the main leader channel.

Recordings with the high-framerate video cameras fielded in the last decade show that lightning leaders, although they involve slightly different space- and time-scales, follow the same pattern as long laboratory sparks.  With integration times of a few microseconds, the observations of \citet{Hill2011/JGRD} for natural stepped leaders and \citet{Biagi2014/JGRD} and \citet{Gamerota2014/GeoRL} for leaders in triggered lightning captured images of the space stem ahead of the leader tip, embedded in a filamentary corona.

Despite these observational advances, our understanding about the physics of stepped leaders is still very incomplete.
Measured optical spectra indicate that the leader temperature reaches around \SI{5000}{K} \citep{Cooray2003/book} for laboratory discharges and up to \SI{30000}{K} in lightning leaders \citep{Orville1968/JGR}, which, in both cases and according to chemical models, suffices to sustain a high ionization \citep{Gallimberti1979/JPhys}.  On the other hand the filaments in the corona, called streamers, are not much above ambient temperature; their ionization, lower than that of leaders, is created mostly at their tips, where they enhance the electric field strongly enough to accelerate electrons up to the threshold of impact ionization \citep{Ebert2010/JGRA}.  Models for the streamer-to-leader transition \citep{Popov2003/PlPhR,da_Silva2013/JGRD} successfully reproduce the transition timescale of around \SI{1}{\micro s} for atmospheric pressure but depend on manually imposing a total electric current that in reality is an outcome of the discharge physics.  They also neglect the longitudinal inhomogeneity of the discharge and therefore they sideline leader stepping and the formation of space stems.  The physical mechanism governing the latter remains a mystery (see e.g. \citep{Biagi2010/JGRD,Bazelyan2010/book});  a recent review \citep{Dwyer2014/PhR} included this problem in the top ten questions in lightning research.

In this letter we show that space stems originate from an attachment instability inside streamer channels.  Since space stems are the key to leader stepping,
our results open the door to the full understanding of this mechanism as well as
its associated radio and energetic particle emissions.  Originally investigated in the 1970s \citep{Douglas-Hamilton1974/JAP,Sigmond1984/JAP}, the attachment instability is triggered by regions of lower conductance per unit length (i.e. conductivity integrated over a cross-section) inside a corona, which we show arise spontaneously when a negative streamer emerges from a leader.  One major and slightly counter-intuitive aspect of our work is that bright regions inside a corona reveal regions of lower, not higher, electron density.  Although this is in complete correspondence with a regular electrical circuit where energy is mostly dissipated in high-resistivity components, this insight has escaped previous interpretations of the space stem.  At high-altitude,  in leader-less discharges (sprites), the attachment instability forms standing patterns called beads and glows \citep{Luque2016/JGRA/temp,Luque2010/GeoRL,Liu2010/GeoRL}.

\section{Model}
Since lightning leaders and long laboratory sparks share the same mechanism of propagation, to simplify our computations we choose to focus here on the propagation of a leader under laboratory conditions.
Typical laboratory leaders span from tens of centimeters to around one meter and are surrounded by streamer coronas with roughly the same extension
\citep{Kostinskiy2018/temp}.  These dimensions are too computationally demanding so, as we detail below, our simulated system is somewhat smaller.

Even then and despite recent progress in three-dimensional streamer simulations 
\citep{Luque2014/NJPh,Teunissen2017/JPhD,Shi2017/JGRD}, a full corona around 
a leader is 
presently out of reach for numerical models.  We opt for simulating a single
streamer that emerges from a leader tip; our assumption here is that the 
surrounding corona is not an essential component of the physics of space stems.
We cannot rigorously justify this assumption but it is beared out by
the similarity between observations and our results.

We thus investigate the formation of space stems ahead of a negative leader channel with a 2D cylindrically symmetric model $(z, r)$ for electric discharges
that includes heating and expansion of the background gas fully self-consistently.  
The background gas follows the equation of state for an ideal gas 
and its dynamics is described by the compressible Euler equations \citep{Popov2003/PlPhR,da_Silva2013/JGRD,Landau1987/fluid}. These are
conservation equations for mass, momentum and energy:

\begin{linenomath*}
\begin{subequations}
\label{eq:euler}
\begin{align}
 & \frac{\partial\rho}{\partial t}+\mathbf{u}\cdot\nabla\rho+\rho\nabla\cdot\mathbf{u}=0,\label{eq:density}\\
 & \frac{\partial\mathbf{u}}{\partial t}+\left(u\cdot\nabla\right)\mathbf{u}+\frac{\nabla p}{\rho}=0,\label{eq:gas_vel}\\
 & \frac{\partial \varepsilon}{\partial t}+\mathbf{u}\cdot\nabla \varepsilon+\frac{p}{\rho}\nabla\cdot\mathbf{u}=\frac{w}{\rho}.\label{eq:internal_energy}
\end{align}
\end{subequations}
\end{linenomath*}

Here $\rho$ is the mass density of air, \textbf{u }is the local velocity
at a given point and time, $p$ is the pressure and $\varepsilon$ is the specific
energy associated to the rotational and translational degrees of freedom, which we assume in thermal equilibrium.
Finally, $w$ is the local dissipated energy from the electric discharge. 
By using equations (\ref{eq:euler}) and the equation of state for an ideal gas, 
we neglect thermal conduction and viscous dissipation, which have little effect on the time-scale of around \SI{100}{ns} 
on which space stems form.

All species are advected along with the fluid with a velocity $ \mathbf{u}$. Furthermore, charged
species drift on top of the background gas motion according to the local value of the electric field $\mathbf{E}$, so the 
resulting velocity is $\mathbf{v_{s}}=\mathbf{u}+\mu_{s}\mathbf{E}$, where $s$ labels the species and $\mu_s$ is the corresponding mobility.
In our model, the dynamics of all charged species is described
by diffusion-drift-reaction equations for electrons and ions, 

\begin{linenomath*}
\begin{equation}
\frac{\partial n_{s}}{\partial t}+\nabla\cdot\left(n_{s}\mathbf{v_{s}}\right)=C_{s}+\nabla\cdot\left(D_{s}\nabla n_{s}\right),\label{eq:number_density}
\end{equation}
\end{linenomath*}
where $n_{s}$ is the number density, $D_{s}$ is the diffusion coefficient
and $C_{s}$ is the net production of species $s$.

The kinetic scheme employed in our simulations includes impact ionization, 
attachment/detachment, and water cluster formation and breaking. 
A detailed description of the scheme can be found in the supplementary material of \citet{Luque2017/PSST}.  The only difference
is that for the three-body attachment reaction 
\begin{linenomath*}
\begin{equation}
  \cee{O2 + O2 + e -> O2- + O2}, \label{eq:Three-body}
\end{equation}
\end{linenomath*}
here we have used the rate from \citet{Kossyi1992/PSST}.

We emphasize the presence of water in the chemical model of our simulations.  The relevance of water vapor for the evolution of streamer channels was previously discussed by \citet{Gallimberti1979/JPhys} and \citet{Luque2017/PSST}.   By clustering around negative ions, even a small quantity of water molecules effectively suppresses electron detachment and thus strongly influences the evolution of the electron density on timescales of tens of nanoseconds.

The electric field $\mathbf{E}=-\nabla\phi$ is
determined by the balance of charged species and satisfies
\begin{linenomath*}
\begin{equation}
-\nabla \cdot \mathbf{E} = \nabla^{2}\phi=-\sum_{s}\frac{q_{s}n_{s}}{\epsilon_{0}},\label{eq:poisson}
\end{equation} 
\end{linenomath*}
where $q_{s}$ is the charge of species $s$ and $\epsilon_{0}$ is the vacuum permittivity, which we assume is also valid for air.

Streamer discharges develop as thin and elongated channels which call for a narrow computational domain.  
To achieve this while suppressing the influence of the radial boundary
conditions in the Poisson’s equation we use the domain decomposition method
described by \citet{Malagon-Romero2018/CoPhC}.  With this method we first find
the electrostatic potential created by the space charges with a homogeneous
Dirichlet boundary at $z=0$ and free conditions in all other boundaries, meaning that the potential decays to zero at large distances.  To the potential obtained in this manner we add a potential $\phi_0 = -E_0 z$ that accounts for an external electric field $E_0$.
The full domain size is $\SI{25}{cm}\times\SI{3}{cm}$.

The term $w$ couples the electrodynamics and bulk gas dynamics and
accounts for dissipated power due to the electric current inside the corona.
But note that this power is distributed
unequally among the degrees of freedom of the underlying gas.  Since the
time-scales involved in the streamer-to-leader transition are too short to
reach thermodynamic equilibrium, the fraction of energy deposited into
different degrees of freedom depends on the local conditions and, in
particular, on
the local electric field \citep{Flitti2009/EPJAP,da_Silva2013/JGRD}.  A small fraction is directly converted into translational energy of gas molecules and quickly  thermalized. A larger amount excites electronic and ionization states; this is responsible for the process of \emph{fast-heating} \citep{Popov2001/PlPhR} and relaxes into thermal energy at timescales on the order of \SI{100}{ns}.  Another fraction of the energy is spent in dissociation of oxygen and nitrogen molecules and, finally, the remaining energy excites vibrational states and its time to thermalization is on the order of one second at ambient temperature and only significant
compared with our relevant timescales once the temperature reaches about \SI{e4}{K}: this relaxation is neglected in the present study.  Since, as we describe later, most heating is due to energies dissipated at or around the conventional breakdown electric field, roughly \SI{30}{kV/cm}, we take the energy branching ratios corresponding to this field, where about half of the energy is frozen into vibrational excitations (see e.g. figure 1 in \citep{Flitti2009/EPJAP}).  Furthermore, since the characteristic time of gas temperature increase is much longer than \SI{100}{ns}, for the sake of simplicity, we consider fast-heating to be instantaneous. Then, we arrive at
\begin{linenomath*}
\begin{equation}
w=\eta\mathbf{j}\cdot\mathbf{E}\label{eq:Joule-heating},
\end{equation}
\end{linenomath*}
where $\eta \approx 0.5$ and $\mathbf{j}=\sum_{s}q_{s}n_{s}\mathbf{v_{s}}$.

Our initial condition consists in a short portion of leader with an 
small ionization patch slightly ahead of the tip that mimicks an irregularity of the leader head.
The initial electron density is thus the sum of an uniform background $n_e^\text{bg}$ plus
\begin{linenomath*}
\begin{subequations}
\begin{equation}
  n_e^{\text{leader}} = n_{e0} \exp\left(-\frac{\max\left(z - z_L ,0\right)^2}{2\sigma_L^2} - \frac{r^2}{2\sigma_L^2}\right),
  \label{initial_leader}
\end{equation}
and
\begin{equation}
  n_e^{\text{seed}} = n_{e0} \exp\left(-\frac{\left(z - z_S\right)^2}{2\sigma_S^2} - \frac{r^2}{2\sigma_S^2}\right),
  \label{initial_streamer}
\end{equation}
\end{subequations}
\end{linenomath*}
where the tip location is $z_L=\SI{5}{cm}$, the seed center is at 
$z_S=\SI{6.1}{cm}$, the $e$-folding lengths are $\sigma_L = \SI{3}{mm}$,
$\sigma_S = \SI{1.5}{mm}$ and the electron density peaks at 
$n_{e0}= \SI{e21}{m^{-3}}$.  The initial electron density is neutralized by an identical density of positive ions.  Note that we selected these initial conditions after a few trials where we disregarded cases in which the streamer branches because these cannot be captured by our cylindrically symmetrical model.  Besides, as mentioned above, to keep our computations feasible, the initial leader is somewhat shorter than experimental stepped leaders.

We have run two different simulations: one with photo-ionization ($n_e^{\text{bg}}=0$) following the method presented by \citet{Luque2007/ApPhL}, and another
with a pre-conditioning of the gas surrounding the leader due to preceding coronas by adding a constant background ionization level $n_e^{\text{bg}}=\SI{e15}{m^{-3}}$. 
In both cases, we observed similar formation of a space stem but
the simulation with photo-ionization exhibited an oscillation of the electric field at the streamer head that we attribute to a numerical artifact due to insufficient resolution for the
smallest length scales involved in photo-ionization \citep{Zhelezniak1982/HTemS,Wormeester2010/JPhD}. Henceforth we limit ourselves to the simulation without photo-ionization.  To check that this does not affect our key results we used another numerical code at our disposal (PESTO, described by \cite{Luque2017/JGRD}) that includes photoionization but does not account for gas heating or long-term chemistry.  Using PESTO, we run simulations with a numerical resolution of \SI{6}{\micro m} that produced results similar to those described below.

The embedding gas is a mixture of 79\% \ce{N2} and 21\% \ce{O2}. Initially, the gas pressure is 1 atm and the mechanical energy is zero. The ambient temperature is 300K and the temperature of the leader follows the same distribution as $n_{e}^{leader}$ with a peak value of 2700K. Note that our model does not include high-temperature chemistry for the leader: in our simulation the role of leader is merely to provide the electrostatic environment for the streamer propagation.
Finally, the simulation is driven by an external electric field pointing towards the leader with magnitude $\vert{E_0}\vert=\SI{10}{kV/cm} + (\SI{20}{kV{cm}^{-1}{\micro s}^{-1}})t$, where $t$ is the simulation time.

With these conditions we simulated the inception and propagation of a streamer emerging from the leader tip.  Our total simulation time was limited to about
\SI{100}{ns} at which point the streamer leaves the simulation domain.  As we see below, this time is enough to see the formation of the space stem but too short to observe the full streamer-to-leader transition.

\section{Results}
\begin{figure*}
  \includegraphics[width=1.0\textwidth]{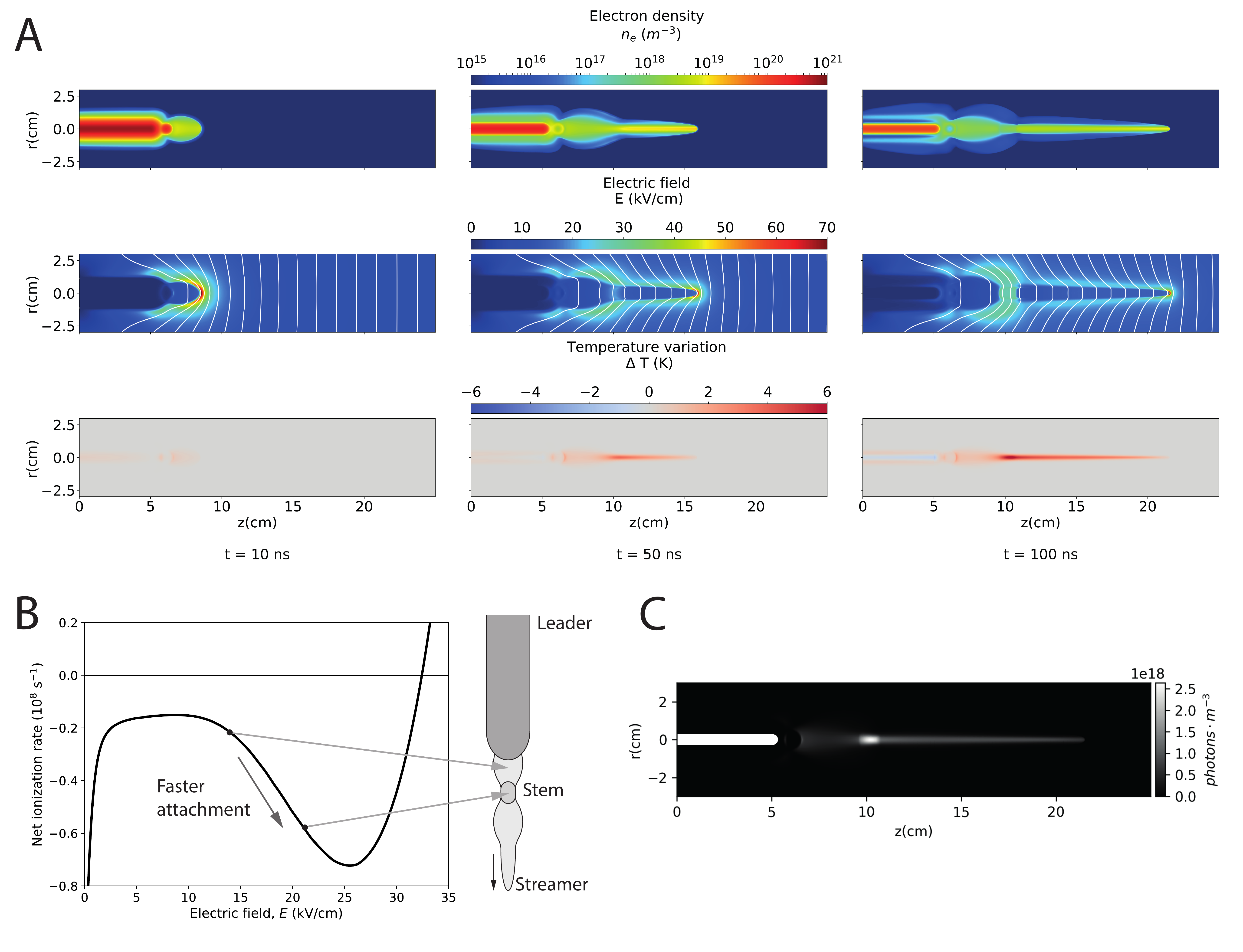}
  \caption{\label{fig:evolution}
    As a streamer propagates out of a leader tip it creates a segment
    of reduced conductivity that evolves into a space stem.  Panel A summarizes the evolution of the streamer in terms of the electron density (top), electric field (middle),  and temperature (bottom). 
The electric field row includes equipotencial lines with constant spacing 12.5 kV, 13.5 kV and 14.5 kV (from left to right). The streamer leaves in its wake a segment of lower conductance per unit length that evolves into a space stem due to the attachment instability process sketched in panel B: a higher electric field accelerates the depletion of electrons, which in turns enhances the electric field.  Finally, in panel C we show that our simulation reproduces the features of a space stem by plotting light emitted in the second positive system of the nitrogen molecule during the full simulation. We have
masked (white region) leader emissions to focus on the space stem.}
\end{figure*}
Figure~\ref{fig:evolution}A summarizes our simulation.  As the streamer emerges from the leader it goes through a narrowing phase where the conductance per unit length decreases.  The charge transport through the streamer channel tends to homogenize the electric current flowing across the streamer channel, which implies a higher electric field in the narrow section.  As sketched in figure~\ref{fig:evolution}B, where we plot the effective ionization rate of air, this enhanced field triggers an attachment instability \citep{Luque2016/JGRA/temp}: the higher field increases the rate of associative electron attachment, decreasing further the conductance per unit length and increasing the field.  This process enhances the electric field inside the narrow section of the channel until it saturates at an electric field where the net ionization curve slopes upward, between 25 and \SI{30}{kV/cm}.  A necessary condition for this process is that the electric field inside the streamer channel steps above the minimum of the effective attachment rate, around \SI{10}{kV/cm}
(see figure \ref{fig:evolution}B).  The emergence of space stems is thus favoured in streamers with high internal electric fields.  


To check that the narrow segment with an enhanced electric field reproduces the observed features of a space stem, we computed the spatial distribution of light emissions.  We included in our model the electron impact excitation of nitrogen molecules to the \NtwoB and \NtwoC electronic states, which are responsible respectively of the first and second positive systems of \ce{N2} (see the Supporting Information \citep{Hagelaar2005/PSST,Capitelli2000/book,Nijdam2014/PSST,LeVeque2002/book,Alghamdi2011/PSP,clawpack,PETSc2016/web,PETSc2016/user} for further details on the chemistry used to describe light emissions).  We found that in our conditions the emissions of light are dominated by the second positive system and panel C of figure~\ref{fig:evolution} shows these emissions integrated over the \SI{100}{ns} of simulation.  There we notice a bright spot embedded in a dim channel, clearly reminiscent of images in high-speed recordings of leader progression \citep{Hill2011/JGRD, Biagi2014/JGRD, Gamerota2014/GeoRL}.  Based on this resemblance we will henceforth use the name \emph{stem} for this bright nucleus within the channel.

Let us now analyze the gas heating produced by the discharge.  This is represented in the bottom row of panel A in figure~\ref{fig:evolution}, where
we show the temperature variation relative to the initial conditions.  The air in the stem heats up about \SI{6}{K} in \SI{100}{ns}. However, in our simulation the electron density decreases both in the stem and in the surrounding channel with a time scale close to \SI{100}{ns}.  This is consistent with previous models and experiments that investigated the effect of the repetition rate in streamer discharges \citep{Nijdam2014/PSST} and therefore it is unlikely that this electron depletion is due to shortcomings of our model.  In our context it implies that the heating ratio diminishes: we do not expect a much higher temperature even if, by increasing our domain size, we extended our simulation time. 

\subsection{Formation of the Space Stem}
Our key result is that the attachment instability is  responsible for locally warmer regions ahead of a leader. A number of processes may reduce the channel conductance per unit length and trigger the instability, among them a jittering of the leader potential during the streamer propagation or pre-existing conductivity or gas-density perturbations along the streamer path \citep{Luque2011/GeoRL,Luque2016/JGRA/temp}.  Neither of these processes was included in our simulations and nevertheless the space stem formed spontaneously, which suggests that isolated stems are robust features of leader propagation.

\begin{figure}
  \includegraphics[width=0.9\columnwidth]{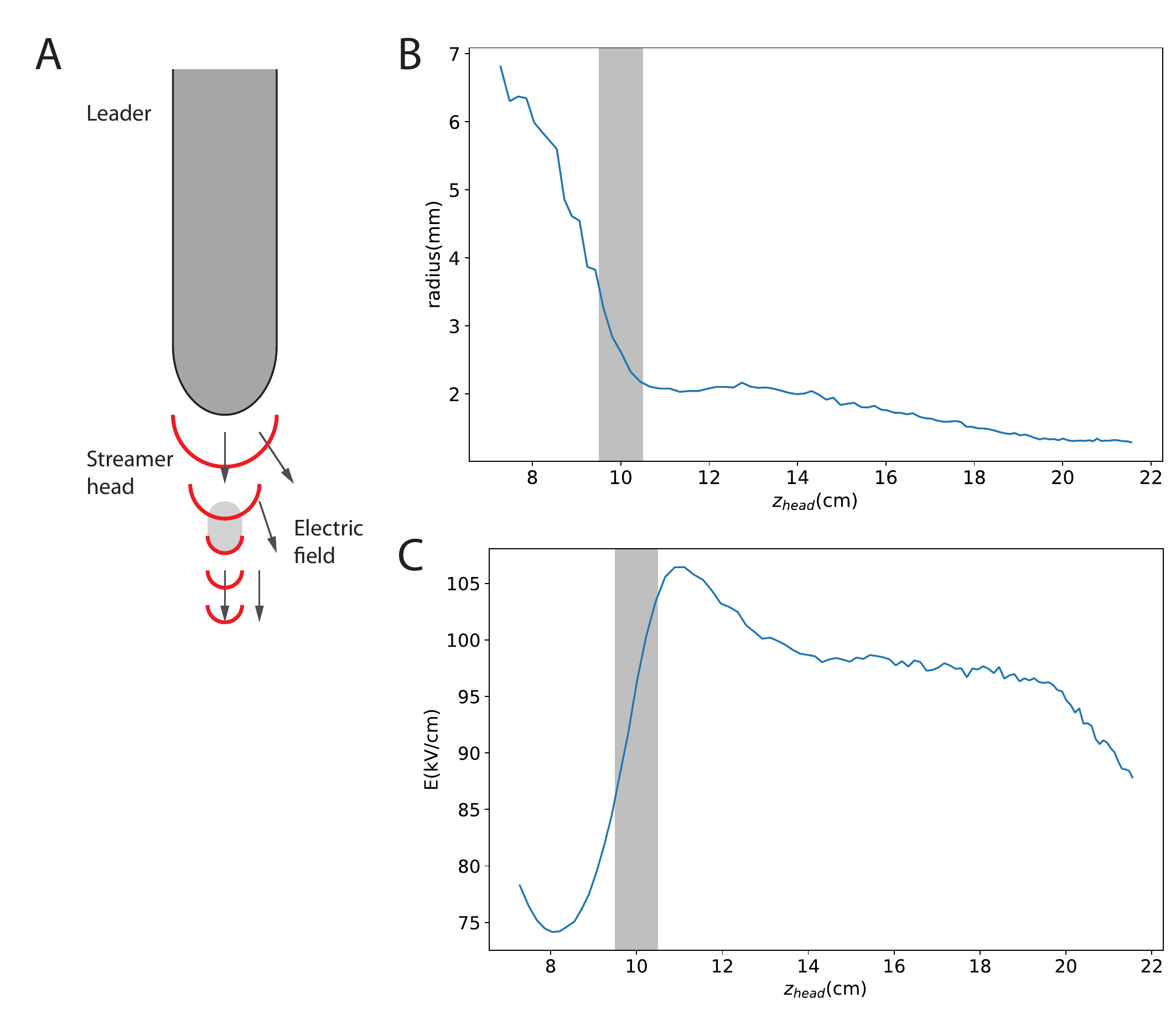}
  \caption{\label{fig:narrowing}
    The space stem emerges due to the narrowing of the streamer channel.  As sketched in panel A, when the streamer is still close to the leader tip it is widened by the diverging electric field lines around the curved leader tip; as it distances itself from the leader, the streamer is driven by a more homogeneous electric field and becomes narrower.  This is shown in panel B, where we plot the streamer radius as a function of time.  The radius is defined here as the radius of curvature on the central axis of the surface defined by the maximum of the electric in the $z$ direction around the head.   The reduction of the radius leads to higher peak electric fields (panel C) and the resulting total channel conductance per unit length exhibits a minimum that afterwards evolves into the space stem as described in figure~\ref{fig:evolution}.}
\end{figure}

\begin{figure}
  \includegraphics[width=0.9\columnwidth]{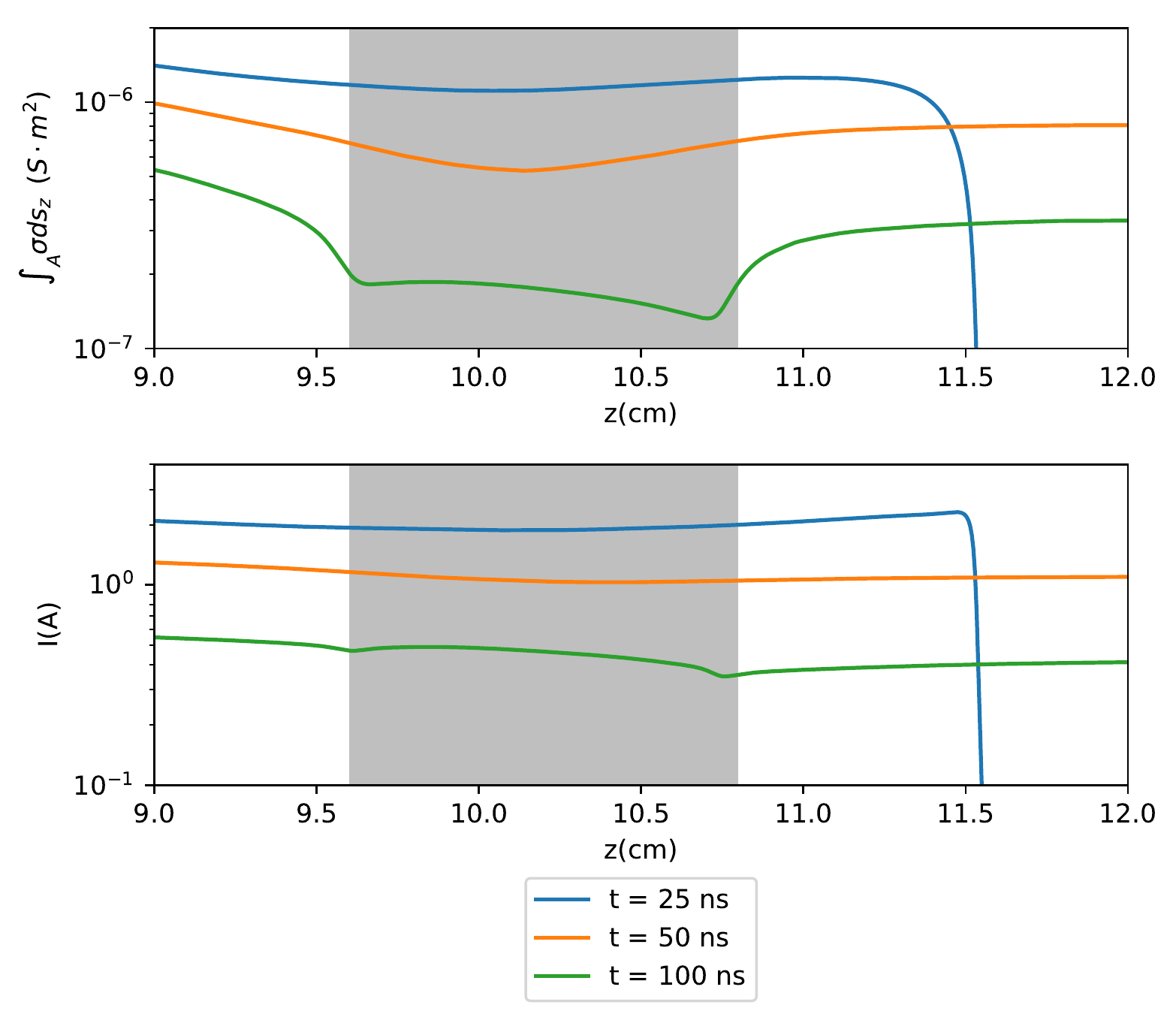}
  \caption{\label{fig:conductance}
    The upper panel shows the conductance per unit length around the space stem (grey area) at 25 ns, 50 ns and 100 ns. As the streamer propagates away from the leader tip (25 ns curve), the channel undergoes
    a narrowing until the conductance per unit length reaches a minimum (space stem). Right after, the channel starts to be able to compensate this narrowing with an 	increase of the electron density produced by a higher electric field and then the conductance per unit length rises. The two remaining curves show latter states of the conductance at the space stem, where
    the electron depletion is clear after attachment instability effects. The lower panel supports the idea that the low conductance in the space stem is countered by a high electric field to achieve an 
  	homogeneous intensity along the channel.}
\end{figure}

In our simulation the stem results from a narrowing of the channel. Note that the narrowing of negative streamers ahead of a leader or a pointed electrode has been observed by \citet{Kochkin2014/JPhD} and by \citet{Kostinskiy2018/temp}.  As we show in figure~\ref{fig:narrowing} the streamer head is initially wide because it is affected by the divergence of electric field lines emerging from the leader's curved tip.  As this divergence decreases away from the leader tip, the streamer head shrinks.  The narrowing of the streamer channel enhances more strongly the electric field at the tip, increasing the degree of ionization left in the streamer head's wake.  The total conductance per unit length of the channel scales approximately as $R^2 n_e$, where $R$ is the channel radius and $n_e$ the electron density: initially the significant decrease of the radius dominates and the conductance per unit length diminishes; afterwards the increase of $n_e$ due to a higher field at the tip overcomes the narrowing and the conductance per unit length increases again.  The resulting minimum is the origin of the space stem as we show in the upper panel of  \ref{fig:conductance}. In the same figure (lower panel) and as we stated before, despite 
the noticeable variation of the conductance per unit length, the intensity is homogeneous across any section of the channel, including the space stem.

Let us now discuss the observed asymmetry between positive and negative leaders.
Stepping is more prominent and readily observable in negative leaders but
there are now clear observations \citep{Kostinskiy2018/temp} that under conditions of high relative humidity, positive leaders also experience stepped progression although space stems have never been observed in positive leaders.
Our results provide a natural explanation for this
asymmetry: the attachment instability is triggered by elevated electric fields
inside a streamer channel and due to stronger ionization in positive streamers, these fields are higher in negative streamers \citep{Luque2008/JPhD} for the same external field.  Besides, positive streamers are initiated more easily
\citep{Liu2012/PhRvL} so they are launched from the leader tip at a lower potential and thus a lower driving electric field than negative streamers.
To check this explanation we run simulations of positive streamers under driving electric fields of \SI{10}{kV/cm} and \SI{7}{kV/cm}; there the attachment instability was triggered only in regions of the channel very close to the leader tip, supporting the idea that steps in positive leaders exist but they are so small that isolated space stems cannot be observed.

\section{Discussion and Conclusions}
Our simulations show that the attachment instability explains the features of space stems ahead of propagating leaders.  However at around 100 ns the overall conductivity of a streamer channel decays, stalling the increase in temperature.  Our results thus stress the role in maintaining the corona played by 
poorly understood processes such as the inception of counter-propagating streamers \citep{Kochkin2016/JPhD,Luque2016/JGRA/temp} or the propagation of successive ionization waves along pre-existing channels \citep{Phelps1974/JATP,Nijdam2014/PSST,Babich2015/JGRA,Rison2016/NatCo}.  Previous models \citep{Popov2003/PlPhR,da_Silva2013/JGRD} missed the relevance of these processes because they were not self-consistent and set a constant current intensity in the channel.  In these models a reduction of the electron density is immediately counteracted by an increase in the applied electric field so electrons are never significantly depleted.  However, no physical mechanism with such an effect has been described in the literature.

As we have already shown,  a single streamer discharge is unable
to dissipate enough power to transit into a leader. Nontheless, the relatively poor conductivity of a streamer corona together with a variable potential at the leader tip imply that there is often a significant electric field within the corona.  This field triggers either new streamer burtsts, as observed by \citet{Kochkin2014/JPhD} or ionization waves retracing previous streamers, as proposed by \citet{Babich2015/JGRA}.  It is also responsible for  counter-streamers seeded by charges in existing stems.  Remarkably, all of these mechanisms have been linked to X-ray emissions from long sparks \citep{Kochkin2015/JPhD,Ostgaard2016/JGRD/c,Ihaddadene2015/GeoRL,Kohn2017/GeoRL,Babich2017/JPhD,Luque2017/JGRD,Babich2015/JGRA} and these X-rays are in turn linked to leader stepping \citep{Dwyer2005/GeoRL/3}.

To check that these mechanisms may indeed explain the streamer-to-leader transition within the currently established observational constraints, we have developed a simplified model of a leader corona that we describe in the Supporting Information. The model shows that ionization waves increasing the  electron density a factor of ten and repeating every 100 ns would lead to a significant increase of the temperature of the channel.  But the main outcome is that a small difference in initial electron density in the stem leads to large differences in the heating rate of this segment compared to the rest of the corona. 

An important simplification of our model is the assumption that a space stem can form within a single streamer channel and that streamer branching, even if present, is not an essential ingredient in the process.  We base this assumption in two key observations: (1) space stems are generally observed as bright segments within longer, dimmer channels \citep{Biagi2010/JGRD,Hill2011/JGRD} and (2) in laboratory images negative streamer coronas contain thick, almost-straight channels with extensions of up to one meter \citep{Kochkin2014/JPhD,Kostinskiy2018/temp}.  Although these channels are surrounded by smaller streamers, there is no reason to believe that these short bifurcations play an essential role in the dynamics of the main channel.
Interestingly, this is not the case for positive coronas \citep{Kochkin2012/JPhD,Kostinskiy2018/temp}, which may be yet another reason for the polarity asymmetry in leader propagation.

Our results explain the formation of brighter and warmer inhomogeneities ahead of a negative leader channel.  This is the first stage in the streamer-to-leader transition in a stepped leader.  The subsequent evolution of the space stem is still not understood: namely we do not know the mechanism that maintains the corona conductivity long enough to reach thousands of degrees.  A full understanding of lightning progression and associated phenomena such as the emission of X-rays will only result from the successful investigation of this mechanism.

\acknowledgments
Information on how to access the code used to run the simulations as well as the output data analyzed in this study is available in the supporting information.
This work was supported by the European Research Council (ERC) under
the European Union H2020 programme/ERC grant agreement 681257. A. Malagón-Romero and A. Luque acknowledge financial support from the State Agency for Research of the Spanish MCIU through the "Center of Excellence Severo Ochoa" award for the Instituto de Astrofísica de Andalucía(SEV-2017-0709). We acknowledge Prof. U. Ebert for useful discussions about the contents of this paper.

\newcommand{\jgr}{J. Geophys. Res.}
\newcommand{\grl}{Geophys. Res. Lett.}
\newcommand{\physrep}{Phys. Rep.}
\newcommand{\mdash}{---}

\end{document}


%
%


\title{\textit{Supporting information} for \\ Spontaneous emergence of space stems ahead of negative leaders in lightning and long sparks}
%
%

\authors{A. Malag\'on-Romero,\altaffilmark{1}
A. Luque\altaffilmark{1}}

\altaffiltext{1}{IAA-CSIC, P.O. Box 3004, 18080 Granada, Spain.}

%
%

%

\begin{article}
\newpage
\linenumbers

%
%

\noindent\textbf{Contents of this file}
\begin{enumerate}
\item Text S1 to S3
\item Figures S1 to S3
\item Table S1 
\end{enumerate}

\noindent\textbf{Introduction}\\
This supporting information provides details of the chemistry accounting for the light emissions (Text S1), a description
of a simplified model for a leader corona (Text S2) that supplements the main scientific conclusions of the paper and a summary about the numerical implementation of the methods used in this work (Text S3). S2 contains three subsections:
a) In S2a we describe the model itself, b) in S2b we set up the model to account for a single corona discharge while in c) S2c, the settings pursue multiple discharges. Table S1 provides the parameter values employed in our simulations. Figures S1 and S2 show the main output of the model for a single and multiple corona discharges respectively.
The data supporting the findings of this study are available at:\\
\url{https://cloud.iaa.csic.es/public.php?service=files&t=4a1b59a05ffed44c475518a39300abac}
The code to obtain the output data that we have analyzed here is available at:\\
 \url{https://gitlab.com/amaro/space_stem.git}

\section*{Text S1. Light Emissions}
Our discharge develops in a high pressure regime (atmospheric pressure),
therefore the dynamics of the charged species is heavily dominated
by collisions.  A fraction of these collisions excites electronic states such
as \NtwoB and \NtwoC. These electronically excited states undergo radiative deactivation:
\begin{linenomath*}
\begin{subequations}
  \label{eq:SPS+FPS}
  \begin{align}
\NtwoB & \rightarrow \gamma_\text{FPS}+\NtwoA,\label{eq:SPS}\\
\NtwoC & \rightarrow \gamma_\text{SPS}+\NtwoB\label{eq:FPS},
\end{align}
\end{subequations}
\end{linenomath*}
and produce emissions known as First Positive System (FPS) and Second Positive System (SPS) respectively. These species can also be
collisionally quenched:
\begin{linenomath*}
\begin{subequations}
\label{eq:quenching}
  \begin{align}
\NtwoB + \cee{N2} & \rightarrow \cee{2N2}, \label{eq:Quen1} \\
\NtwoB + \cee{O2} & \rightarrow \cee{N2 + 2O},\label{eq:Quen2} \\
\NtwoC + \cee{N2} & \rightarrow \Ntwoaprime + \cee{N2}, \label{eq:Quen3} \\
\NtwoC+ \cee{O2} & \rightarrow \cee{N2 + O(3P) + O(1S)}.\label{eq:Quen4}
\end{align}
\end{subequations}
\end{linenomath*}
In our code we used electron impact excitation rates for reactions (\ref{eq:SPS+FPS}) obtained from BOLSIG+ \citet{Hagelaar2005/PSST} using
the cross-section database \citet{Phelps/Online}.  The reaction rates for the collisional quenching reactions (\ref{eq:quenching}) have been obtained from \citet{Capitelli2000/book}.

\section*{Text S2. Two-Sphere Model for a Leader Corona}
In the main text we described in detail the initial \SI{100}{ns} of the streamer-to-leader transition.  This time is sufficient to establish the space stem but it is clearly too short to reach the characteristic temperatures inside a leader.
Unfortunately, for the reasons explained in the main text, currently microscopical models cannot be extended to long times.  However we may obtain a glimpse of the most important physics at these longer timescales.

We present here a streamlined and heavily simplified model of the corona
ahead of a leader.  Although it is clearly insufficient to produce accurate predictions, it illustrates the physics of the streamer-to-leader transition shows two things: (1) that for the streamer-to-leader transition to occur there must be a mechanism that either creates new streamers or increases the conductivity of existing channels by means of consecutive ionization waves and (2) that a small difference in initial electron density in the stem leads to large differences in the heating rate of this segment compared to the rest of the corona.

\subsection*{Text S2a. Model Description}
\begin{figure*}
  \includegraphics[width=\textwidth]{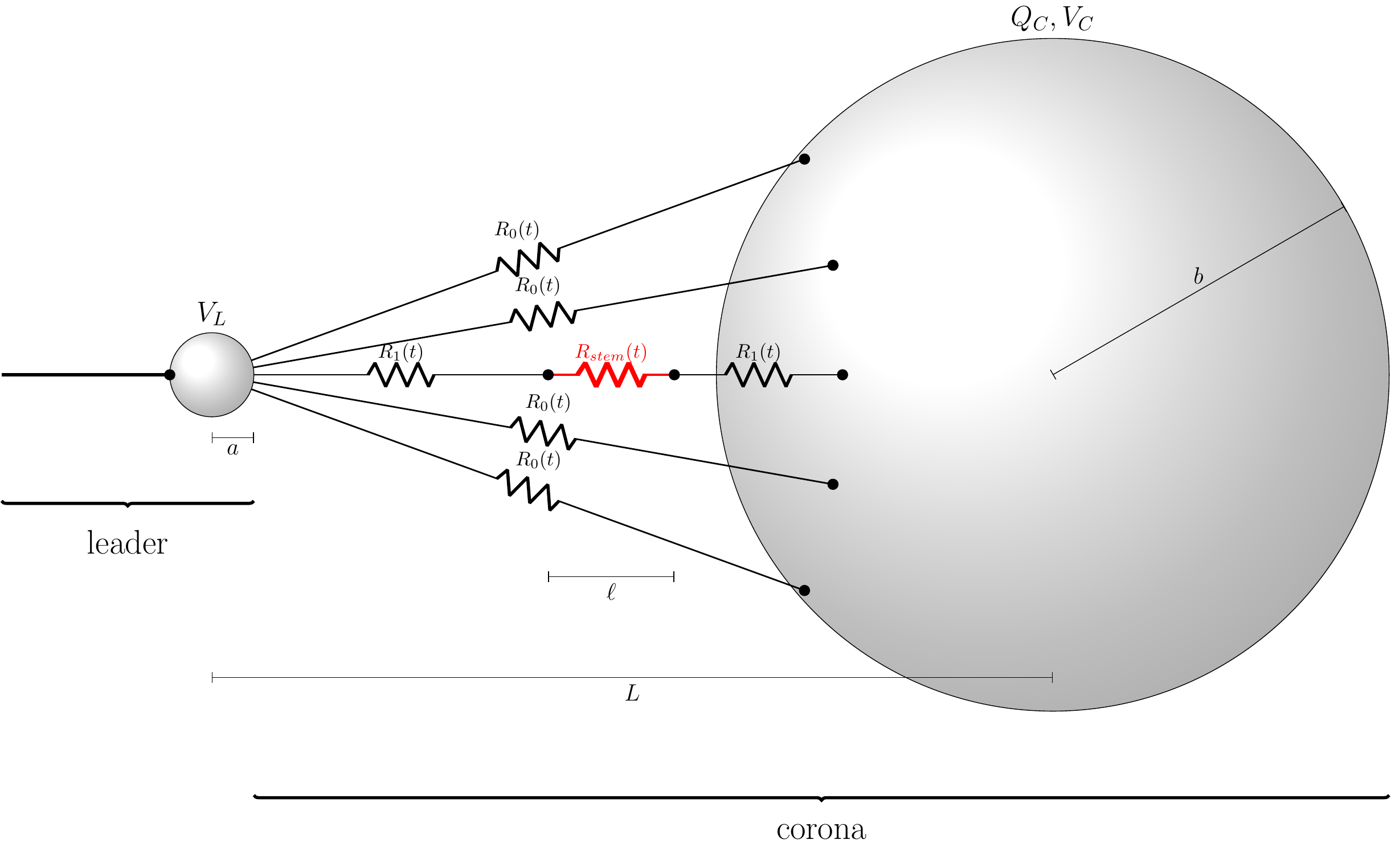}
  \caption{\label{fig:scheme}
    Sketch of the model.  Two conducting spheres, representing respectively the
    leader tip and the corona space charge are connected by channels representing
    the streamer discharges.  Most of these channels are uniform and modeled
    as single resistors. One of the channels is perturbed to contain a space
    stem, represented by a different resistor (shown in red).
  }
\end{figure*}

The model's geometry is sketched in figure~\ref{fig:scheme}.  The leader tip is
mimicked by a conducting sphere of radius $a$ at a potential $V_L$,
whereas the electrical charge in the corona is distributed within another
sphere with radius $b$.  The centers of the two spheres are separated
by a length $L$.  In order to simplify our calculations we assume that the
corona sphere has a uniform potential $V_C$ arising from a total corona
charge $Q_C$.  The electrostatic system of two conducting spheres with potentials $V_L$, $V_C$ and charges $Q_L$ and $Q_C$ is defined by a capacitance matrix
$C$ such that
\begin{linenomath*}
\begin{equation}
  \label{capacitance}
  \left(
  \begin{array}{c}
    Q_L \\ Q_C
  \end{array}\right) = \left(
    \begin{array}{cc}
      C_{LL} & C_{LC} \\
      C_{CL} & C_{CC}
  \end{array} \right) \left(
  \begin{array}{c}
    V_L \\ V_C
  \end{array}\right),
\end{equation}
\end{linenomath*}
where the elements of $C$ can be calculated by repeated application of the
method of images.  From (\ref{capacitance}) we obtain
\begin{linenomath*}
\begin{equation}
  V_C = -\frac{C_{CL}}{C_{CC}} V_L + \frac{Q_C}{C_{CC}}.
\end{equation}
\end{linenomath*}

The two spheres are connected by streamer channels, which this model represents
as linear electrical resistors.  Of these, $N$ are identical, unperturbed
resistors of length $L$ (we neglect differences in these distances) whereas
one perturbed channel contains the space stem and is thus divided into three
serially connected resistors: the space stem, of length $\ell$ is surrounded by
two channels of length $(L-\ell)/2$.

The current inside each resistor is generated by the drift of current carriers, each with mobility $\mu_s$, where $s=1,\dots$ indexes the species (this includes electrons and positive and negative ions).  The underlying charge carrier densities and mobilities in the streamers are respectively $n_{0,s}$ in the unperturbed, long channels, $n_{\text{stem},s}$ in the stem and $n_{1,s}$ in the
two channel segments surrounding the stem.  These densities are uniform within the channels, all of them of cylindrical shape and with a radius $r$.

Assuming now that the electric field that drives the charge carriers inside each channel can be approximated by its average, we obtain the current in the unperturbed channels as
\begin{linenomath*}
\begin{equation}
  I_0 = \frac{\pi r^2 e (V_L-V_C)}{L} \sum_s \mu_s n_{0,s} = \frac{V_L-V_C}{R_0},
\end{equation}
\end{linenomath*}
where the resistance $R_0$ is defined as
\begin{linenomath*}
\begin{equation}
  R_0 = \frac{L}{\pi r^2 e} \left(\sum_s \mu_s n_{0,s}\right)^{-1}.
\end{equation}
\end{linenomath*}
Defining similarly the resistance of the stem as
\begin{linenomath*}
\begin{equation}
  R_\text{stem} = \frac{\ell}{\pi r^2 e} \left(\sum_s \mu_s n_{\text{stem},s}\right)^{-1},
\end{equation}
\end{linenomath*}
and that of the two channel surrounding the stem as
\begin{linenomath*}
\begin{equation}
  R_1 = \frac{(L-\ell)}{2\pi r^2 e}
  \left(\sum_s \mu_s n_{1,s}\right)^{-1},
\end{equation}
\end{linenomath*}
yields for the current in the perturbed channel
\begin{linenomath*}
\begin{equation}
  I_1 = \frac{V_L-V_C}{R_\text{stem} + 2R_1}.
\end{equation}
\end{linenomath*}
The charge accumulation in the corona then follows
\begin{linenomath*}
\begin{equation}
  \od{Q_C}{t} = NI_0 + I_1.
\end{equation}
\end{linenomath*}

The species densities evolve according to the chemical system described in the main text subjected to electric fields averaged over the extension of each path, which can be calculated from Ohm's law as
$\bar{E_c} = R_c I_c/l_c$, where $c$ indicates the kind of channel (unperturbed, stem or stem-neighbors) and $l_c$ is the channel's length. Furthermore, each channel dissipates energy at a rate $R_c I_c^2$ and therefore its temperature increases at a rate
\begin{linenomath*}
\begin{equation}
  \od{T}{t} = \frac{\eta R_C I_C^2}{\pi r^2 l_c n_{\text{air}} c_V},
\end{equation}
\end{linenomath*}
where $n_\text{air}$ is the number density of air at standard temperature and pressure and $c_V$ is the specific heat capacity of air, which we take as $c_V=(5/2) k$, $k$ being Boltzmann's constant.  As in the main text, $\eta$ stands for the fraction of energy deposited into thermalized degrees of freedom; we take here too $\eta=1/2$.

Note that as the stem heats up some processes that are not included in our chemical model, such as vibrational-translational relaxation, play an increasingly significant role.  Therefore high temperatures in this simplified model cannot be considered as quantitative predictions.  

\begin{table}
\begin{tabular}{ll}
Parameter & Value \\ \hline
$L$ & \SI{1}{m} \\
$a$ & \SI{1}{cm} \\
$b$ & \SI{25}{cm} \\
$r$ & \SI{0.5}{mm} \\
$V_L$ & \SI{1}{MV} \\
$N$ & 50 \\
\end{tabular}
\caption{Parameter values employed in our simulations.\label{tbl:params}}
\end{table}

The parameters used in the following simulations are listed in 
table~\ref{tbl:params}.  As initial conditions we set an electron density $n_e=\SI{e19}{m^{-3}}$, balanced by \ce{N2+} and \ce{O2+} in a ratio matching the air fractions of molecular nitrogen and oxygen.  In the ``stem'' resistor this initial density is reduced by a factor 0.75.  Once this different initial condition is set, all resistor follow the same evolution equations.  Note that our input values have not been fine-tuned to obtain the results described below.

\subsection*{Text S2b. Single Corona Discharge}
\begin{figure}
  \includegraphics[width=0.7\columnwidth]{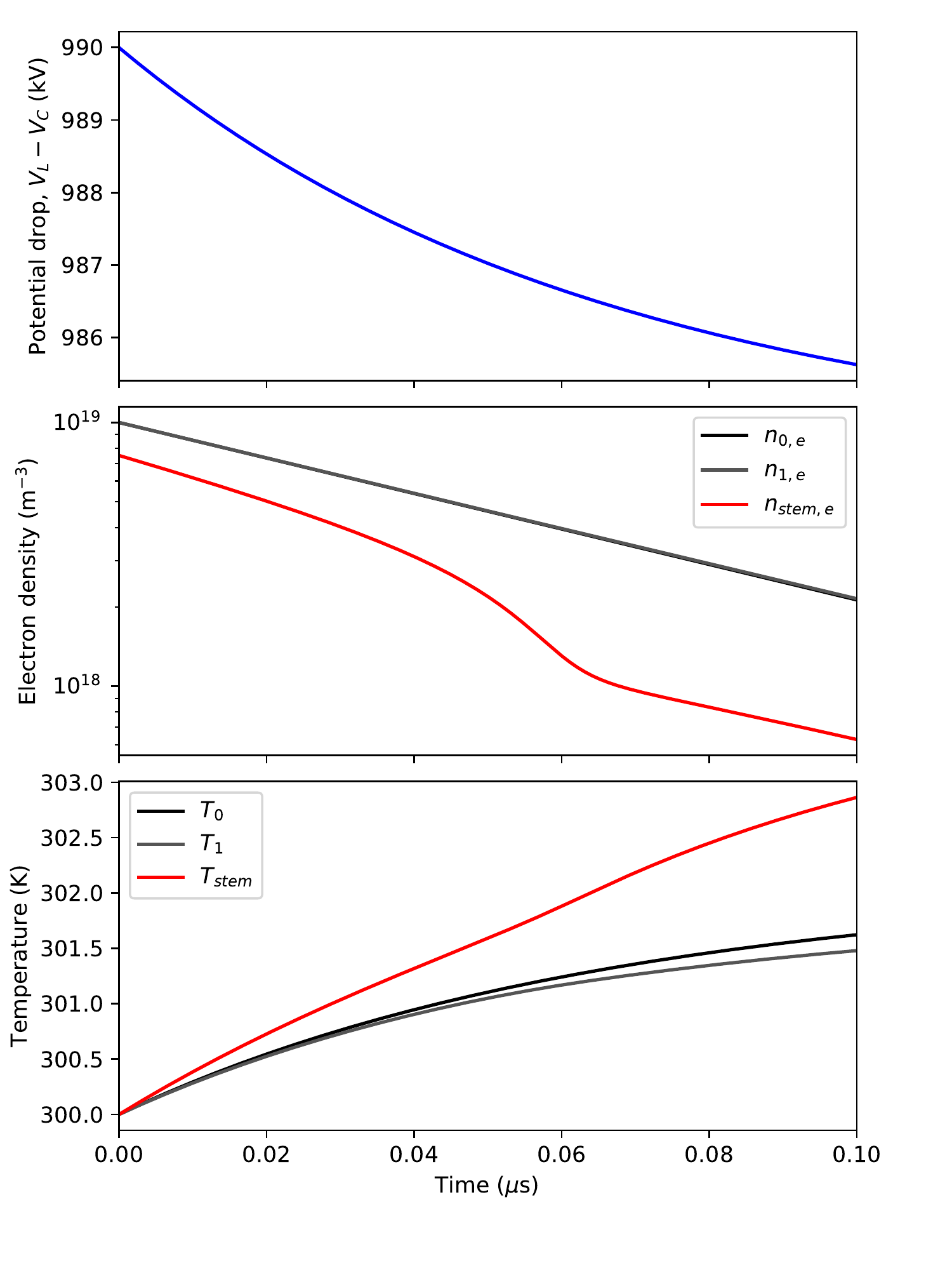}
  \caption{\label{fig:single_discharge} Simulation of a single corona discharge in the two-sphere model. Here we show \SI{100}{ns} of simulation.  The top panel shows a small decrease in the potential drop; the central panel shows that the electron density decays quickly, faster in the space stem that in the other channels.  Finally, the lower panel plots a small increase of gas temperature, which is nevertheless much more significative in the stem.
}
\end{figure}

As a first step, let us check that results from this simplified model are broadly consistent with the fluid model described in the main text.  Figure~\ref{fig:single_discharge} shows the evolution of the most relevant variables of the model.  The main result is that, as we noticed in the main text, the electron density is depleted with a timescale of around \SI{100}{ns}.  Hence the temperature in any of the channels does not increase further than a few Kelvin.  However, the small perturbation in electron density introduced in the ``stem'' channel is sufficient to excite the attachment instability described in the paper and thus drives this component to higher electric fields and more dissipation.  The increase in temperature is thus significantly higher in this segment but still far below that needed to transition to a leader.

Note however that the potential drop between the leader and the corona has barely bulged and there is still a high potential in the leader that is available to initiate new discharges.

\subsection*{Text S2c. Multiple Discharges}
\begin{figure}
  \includegraphics[width=0.7\columnwidth]{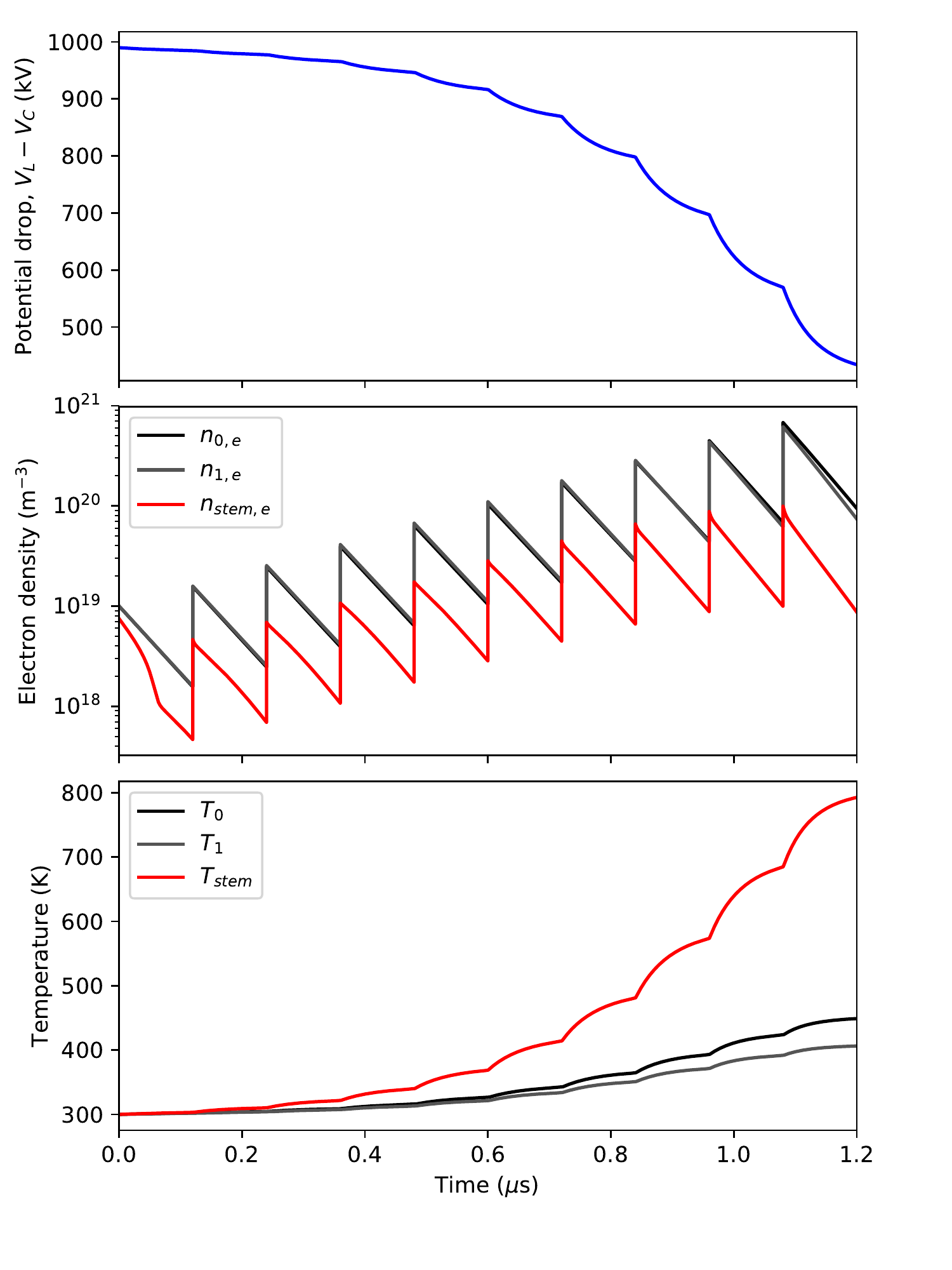}
  \caption{\label{fig:multiple_discharge} Corona discharge followed by multiple ionization waves that sustain a high ionization in the streamer channels.  In this case the electrostatic potential drop (top panel) decays significatively, whereas the electron density (middle panel) increases slowly.  The temperature of the channels (lower panel) increases up to around \SI{800}{K} for the stem and only to around~\SI{400}{K} for the rest of the channels.
  }
\end{figure}
We hypothesize that the large electrostatic potential remaining at the leader tip after the electron density has been depleted generates subsequent ionization waves that prevent the conductivity region ahead of the leader tip to disappear completely.  To substantiate this hypothesis we provide here an example of how the physics of the streamer-to-leader transition may work on long timescales after the space stem has formed.

Assume then that due to its high potential, the leader tip launches successive streamer-like ionization waves that propagate along or close to previous existing channels.  A similar process was observed by \citet{Nijdam2014/PSST}.  As an example, let us assume that these waves are launched every \SI{100}{ns} and that their effect is to increase the channel ionization by a factor 10.  Given their fast timescales, these ionization waves can be implemented in our model by instantaneous increases of the electron and ion densities.

The result is plotted in figure~\ref{fig:multiple_discharge}, where we show the evolution of the system after 10 ionization waves.  The gas heating in this case is stronger, with the stem reaching a temperature close to \SI{800}{K} compared to only about \SI{400}{K} for the rest of the corona channels.

\section*{Text S3. Numerical Implementation}
In the form used in this work, the compressible Euler equations are
hyperbolic equations with additional source terms. Finite Volume
Methods are suitable to solve these equations once the source terms
are properly treated. To solve these equations we have used CLAWPACK/PETCLAW \citep{LeVeque2002/book,Alghamdi2011/PSP,clawpack}.
 PETCLAW is built upon PETSc \citep{PETSc2016/web,PETSc2016/user} and allows us to split the simulation
domain into different subdomains (problems) that can be solved in parallel.
The Poisson's equation is solved using the Improved Stabilized version
of BiConjugate Gradient solver from the PETSc numerical library.
The space resolution was \SI{25}{\micro m} and we used an adaptive time-step constrained by a Courant-Friedrichs-Levy number of 0.005 and by the shortest chemical time-scale $\tau_s = n_s (dn_s/dt)^{-1}$ among all species $s$.

\newcommand{\newblock}{}

\end{article}